\documentclass[twocolumn,aps,prl,amsmath,showpacs]{revtex4}
\usepackage{graphicx}
\def\be{\begin{equation}}
\def\ee{\end{equation}}
\def\bc{\begin{center}}
\def\ec{\end{center}}
\def\lan{\langle}
\def\ran{\rangle}
\def\r{{\vec r}}
\def\k{{\vec k}}
\begin{document}

\title{Static and dynamic heterogeneities in a model for irreversible gelation}
\author{T. Abete, A. de Candia, E. Del Gado,
A. Fierro, and A. Coniglio}  \affiliation{ Dipartimento di Scienze
Fisiche, Universit\`{a} degli Studi di Napoli ``Federico II'', INFN,
CNR-INFM  Coherentia and CNISM, via Cinthia, 80126 Napoli, Italy}

\date{\today}

\begin{abstract}
We study the structure and the dynamics in the formation of
irreversible gels by means of molecular dynamics simulation of a
model system where the gelation transition is due to the random
percolation of permanent bonds between neighboring particles. We
analyze the heterogeneities of the dynamics in terms of the
fluctuations of the intermediate scattering functions: In the sol phase
close to the percolation threshold, we find that this dynamical
susceptibility increases with the time until it reaches a plateau.
At the gelation threshold this plateau scales as a function of the
wave vector $k$ as $k^{\eta -2}$, with $\eta$ being related to the
decay of the percolation pair connectedness function. At the lowest
wave vector, approaching the gelation threshold it diverges with the
same exponent $\gamma$ as the mean cluster size.
These findings suggest an alternative way of measuring critical exponents
in a system undergoing  chemical gelation.
\end{abstract}

\pacs{82.70.Gg, 61.25.Hq, 64.60.Ak} \maketitle In spite of the
relevance of sol-gel processes in polymer physics, a comprehensive
understanding of dynamics in irreversible gelation has not been
achieved yet. Following the pioneering work by Flory and de Gennes
\cite{flo}, it is generally accepted that the divergence of the
cluster size and the formation of a percolating network of permanent
bonds is responsible for the sol-gel phase transition, with a
critical increase of the viscosity coefficient and the onset of an
elastic response close to the gelation transition.
The question of how the cluster size can be related to a
thermodynamic measurable quantity like the fluctuation of the order
parameter near a critical point, is still, to our knowledge, an open
question. A second question of what are the fundamental analogies
and differences between the complex dynamics of polymer gels,
colloidal gels, structural glasses and spin glasses has occasionally
been discussed in the literature \cite{adam,gel_dyn_theo}, but never
fully clarified.

In this paper we address these two questions by means of theoretical
arguments and molecular dynamics (MD) simulations. We first study by
MD the gel formation in a model system, where neighboring particles
(monomers) are linked by permanent bonds to form clusters of
different sizes. By varying the volume fraction $\phi$ the system
exhibits a percolation transition at $\phi_c$, in the same
universality class as random percolation. We analyze the dynamics in
the sol phase, by means of the self intermediate scattering
functions and show that the percolation transition coincides with a
dynamical transition characterized first by stretched exponentials
and at the percolation threshold by a power law behavior, as found
in the experiments \cite{gel_dyn_exp} and in some recent numerical
works on the lattice \cite{cubetti}. To compare chemical gelation
with the slow dynamics observed in colloidal gels or other
disordered systems, such as glasses and spin-glasses, we measure the
dynamic susceptibility defined as the fluctuations of the self
intermediate scattering function. In supercooled liquids this or
similar quantities have been introduced to characterize the behavior
of dynamical heterogeneities \cite{franz, biroli}, which typically
grow with time, reach a maximum, and then decrease at large time.
This behavior, related to the growth of a dynamical correlation
length, is a consequence of the transient nature of dynamic
heterogeneities. In this model for chemical gelation, instead, we
find that, approaching the gelation threshold in the sol phase, the
dynamic susceptibility increases with time, until it reaches a
plateau in the long time limit. This behavior is due to the presence
of static heterogeneities (clusters), which being persistent do not
lead to the decay of the non linear susceptibility.
We argue in fact that the non linear dynamical susceptibility, in
the infinite time limit, $\chi_{as}(k,\phi)$ for  $k\to 0$,
coincides with the mean cluster size. The numerical data strongly
support this result and show that, for small $k$, it obeys the
following scaling behavior
$\chi_{as}(k,\phi) = k^{\eta-2}f(k\xi)$, where $\xi \sim(\phi_c
-\phi)^{-\nu}$ is the connectedness length (the linear size of a
critical cluster) and $\nu$ the associated critical exponent, while
$2-\eta= \gamma /\nu$, $\gamma$ being the mean cluster size
exponent \cite{stauffer}.
These relations link the
mean cluster size to the fluctuations of the intermediate scattering
function, which  can be measured from light scattering experiments
\cite{self}. Therefore our results can be tested experimentally
and offer a new alternative to previous methods \cite{adconst},
to measure percolation exponents in a sol-gel transition.

{\em The model - } We consider a $3d$ system of $N= 1000$ particles
interacting via a Lennard-Jones potential, truncated in order to have
only the repulsive part:
$$
U_{ij}^{LJ}=\left\{ \begin{array}{ll}
4\epsilon[(\sigma/r_{ij})^{12}-(\sigma/r_{ij})^6+\frac{1}{4}], & r_{ij}<2^{1/6}\sigma \\
0, & r_{ij}\ge2^{1/6}\sigma \end{array} \right.
$$
where $r_{ij}$ is the distance between the particles $i$ and $j$.
After a first equilibration, we introduce quenched bonds between particles
whose relative distance is smaller than $R_0$ by adding
an attractive potential:
$$
U_{ij}^{FENE}=\left\{ \begin{array}{ll}
-0.5 k_0 R_0^2 \ln[1-(r_{ij}/R_0)^2], & r_{ij}< R_0\\
\infty, & r_{ij}\ge R_0 \end{array} \right.
$$
representing a finitely extendable nonlinear elastic
(FENE)\cite{FENE}. The system is then further thermalized. We have
chosen $k_0=30\epsilon/\sigma^2$ and $R_0=1.5\sigma$ as in Ref.
\cite{FENE} and performed MD simulations in a box of linear size $L$
(in units of $\sigma$) with periodic boundary conditions. The
equations of motion were solved in the canonical ensemble (with a
Nos\'e-Hoover thermostat) using the velocity-Verlet algorithm
\cite{Nose-Hoover} with a time step $\Delta t=0.001\delta\tau$,
where $\delta\tau=\sigma(m/\epsilon)^{1/2}$, with $m$ the mass of
particle. In our reduced units the unit length is $\sigma$, the unit
energy $\epsilon$ and the Boltzmann constant $k_B$ is set equal to
$1$. The temperature is fixed at $T=2$ and the volume fraction
$\phi=\pi\sigma^3N/6L^3$
is varied from $\phi=0.02$ to $\phi=0.2$.
By varying the volume fraction we find
that the system undergoes a random percolation transition. From a
standard finite size scaling analysis \cite{stauffer}, we have
obtained the percolation threshold $\phi_c$, and the critical
exponents $\nu$ (which governs the power law divergence of the
connectedness length $\xi\sim|\phi-\phi_c|^{-\nu}$ as the critical
point is approached from below) and $\gamma$ (governing the power
law divergence of the mean cluster size $\chi\sim
|\phi-\phi_c|^{-\gamma}$). The results obtained are
$\phi_c=0.10\pm0.02$, with critical exponents $\nu=0.88\pm0.05$ and
$\gamma=1.8\pm0.1$ in agreement with random percolation.

\begin{figure}
\begin{center}
\includegraphics[width=8cm]{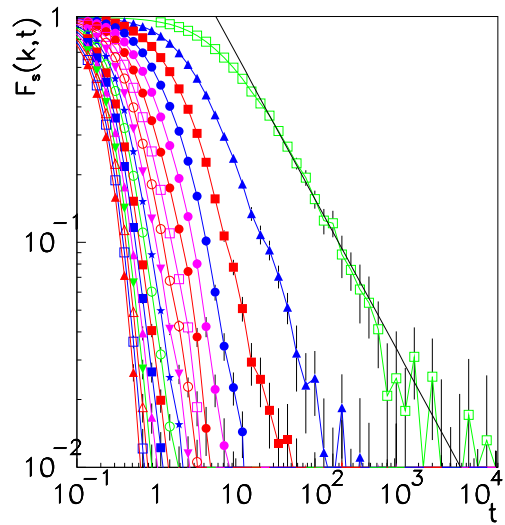}
\caption{
$F_s(k,t)$ for decreasing values of $k$ from left to right at
$\phi= 0.09$ as a function of the time $t$. At $k=k_{min}\simeq0.35$
the data are fitted by a power law $\sim t^{-c}$
with  $c=0.65\pm0.03$ (straight line). } \label{fig4_1}
\end{center}
\vspace{-0.5cm}
\end{figure}

\begin{figure}
\begin{center}
\includegraphics[width=8cm]{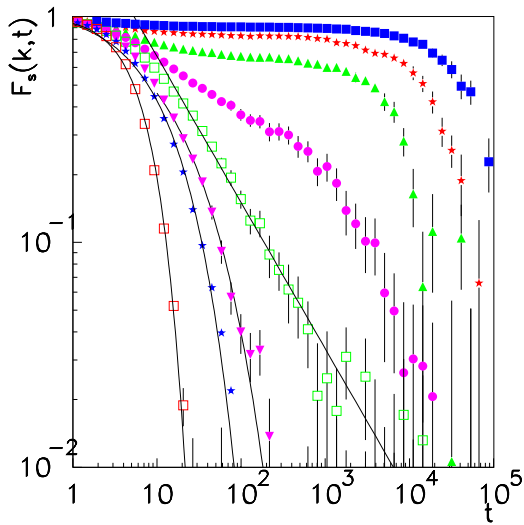}
\caption{$F_s(k,t)$ for $\phi= 0.02$, $0.07$, $0.09$, $0.10$,
$0.11$, $0.12$, $0.13$ (from left to right), and $k=k_{min}$ as a
function of $t$. The full lines are, from left to right, an
exponential, a stretched ($\beta=0.75\pm0.01$ and
$\beta=0.58\pm0.01$) and a a power law $\sim t^{-c}$ with
$c=0.65\pm0.03$ (straight line).} \label{fig4}
\end{center}
\vspace{-0.5cm}
\end{figure}

{\em Dynamical properties -} The dynamics at equilibrium is analyzed
by measuring the self intermediate scattering function $F_s(k,t)$
defined as  $F_s(k,t)=\left[\lan \Phi_s(k,t)\ran\right]$ where
$\Phi_s(k,t)=\frac{1}{N}\sum_{i=1}^N e^{i\vec{k}\cdot(\vec{r}_i(t)-
\vec{r}_i(0))} $, $\langle \dots \rangle$ is the thermal average for
a fixed bond configuration and $[\dots]$ is the average over the bond
configurations. In our simulations the average is over $30$
independent bond configurations.
At low volume fractions $F_s(k,t)$ decays to zero
following an exponential behavior for all the wave vectors $k$
considered. Increasing the volume fraction, close to the percolation
threshold, at low wave vectors the long time decay starts to follow
a stretched exponential behavior $\sim e^{-{(t/\tau)}^\beta}$: The
cluster size distribution has already started to widen and
therefore, over sufficiently large length scales (small $k$), the
behavior of $F_s(k,t)$ is due to the contribution of different
clusters, characterized by different relaxation times, whose
superposition produces a detectable deviation from an exponential
law. Close to $\phi_c$ (see  Fig. \ref{fig4_1}) the onset of a power
law decay is observed at the lowest wave vector, $k_{min}=2\pi/L$, indicating a
critical slowing down due to the onset of a percolating cluster \cite{daoud}.
The behavior of $F_s(k,t)$ for $k=k_{min}$
(plotted in Fig. \ref{fig4} as a
function of the time for different volume fractions $\phi$)
gives the relaxation dynamics over length scales of the
order of the system size.
As $\phi$ increases towards $\phi_c$, we observe a crossover from an
exponential decay to a stretched exponential one, with $\beta$
decreasing as a function of the volume fraction. At $\phi_c$ the
long time decay displays a power law behavior $\sim t^{-c}$. If
$\phi$ increases further, the system is out of equilibrium.
For long values of the waiting time, we find
that $c$ decreases until the long time decay
becomes indistinguishable from a logarithm behavior and eventually a
two step decay appears.
These dynamical features well reproduce the experimental
observations in different systems close to the gel transition
\cite{gel_dyn_exp} and agree with previous numerical results
obtained on a lattice model \cite{cubetti}.

\begin{figure}
\begin{center}
\includegraphics[width=8cm]{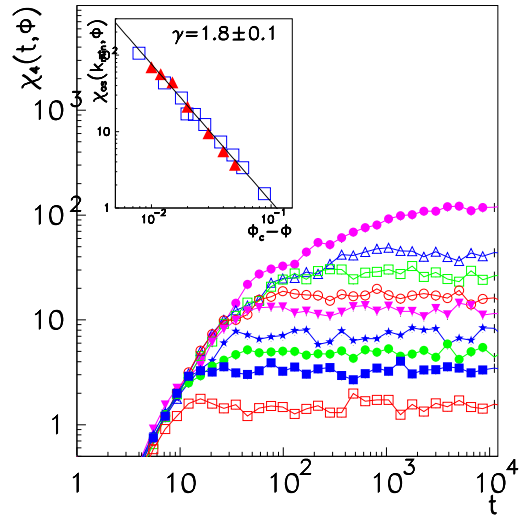}
\caption{{\bf Main frame}: $\chi_4(k_{min},t)$ as a function of $t$ for
$\phi= 0.02$, $0.05$, $0.06$, $0.07$, $0.08$, $0.085$,
$0.09$, $0.095$, $0.10$ (from bottom to top). {\bf Inset}:
Asymptotic values of the susceptibility (full triangles),
$\chi_{as}(k_{min},\phi)$ and mean cluster size (open
squares) as a function of $(\phi_c-\phi)$.
The data are fitted by the power law
$(\phi_c-\phi)^{-\gamma}$ with $\gamma=1.8\pm 0.1$.}
\label{fig5}
\end{center}
\vspace{-0.5cm}
\end{figure}

We now analyze and discuss the behavior of the dynamical
susceptibility $\chi_4(k,t)$ associated to the fluctuations of
$\Phi_s(k,t)$, i.e. $\chi_4(k,t)=N\left[\rule{0pt}{10pt}\lan
|\Phi_s(k,t)|^2\ran-\lan \Phi_s(k,t)\ran^2\right]$.
$\chi_4(k_{min},t)$ is plotted in Fig. \ref{fig5} for $\phi \le
\phi_c$: Differently from the non monotonic behavior typically
observed in supercooled glassy systems, we find that
$\chi_4(k_{min},t)$ increases with time until it reaches a plateau,
whose value increases as a function of $\phi$.

Indeed we argue that in the thermodynamic limit, when $k \to 0$ and
$t \to \infty$, $\chi_4(k,t)$ tends to the mean cluster size.
Being
$\lim_{t\to\infty}\lan\Phi_s(k,t)\ran=0$, we have in the long time
limit
\begin{eqnarray}
\lim_{t\to\infty} \chi_4(k,t) &=& \lim_{t\to\infty}
\frac{1}{N}\sum_{i,j=1}^N\,\left[ \lan
e^{i\k\cdot(\r_i(t)-\r_i(0))}e^{-i\k\cdot(\r_j(t)-\r_j(0))}\ran\right]
\nonumber\\
&=&\frac{1}{N}\sum_{i,j=1}^N\,\left[ \left|\lan
e^{-i\k\cdot(\r_i-\r_j)}\ran\right|^2\right] \label{self1}
\end{eqnarray}
In Eq.\ (\ref{self1}) we have considered
that, for large enough time $t$, the term
$e^{-i\k\cdot(\r_i(t)-\r_j(t))}$ is statistically independent from
$e^{-i\k\cdot(\r_i(0)-\r_j(0))}$, so that we can factorize the
thermal average. In the last term of Eq.\ (\ref{self1}), we may now
separate the sum over connected pairs ($\gamma_{ij}=1$, that is
pairs belonging to the same cluster), and disconnected ones
($\gamma_{ij}=0$, that is pairs belonging to different clusters), so
that
\begin{eqnarray}
\lim_{t\to\infty}\chi_4(k,t)&=&\frac{1}{N}\sum_{i,j=1}^N\, \left[
\gamma_{ij}\left|\lan e^{-i\k\cdot(\r_i-\r_j)}\ran\right|^2\right]
\nonumber\\
&+&\frac{1}{N}\sum_{i,j=1}^N\,\left[ (1-\gamma_{ij})\left|\lan
e^{-i\k\cdot(\r_i-\r_j)}\ran\right|^2\right] \label{chi4sum}
\end{eqnarray}

For $\phi<\phi_c$, clusters will have at most a linear size of order
$\xi$, so that the relative distance $|\r_i-\r_j|$ of connected
particles will be $\le\xi$. Therefore, for $|\k|\ll\xi^{-1}$ and
$\gamma_{ij}=1$, we have $\lan e^{-i\k\cdot(\r_i-\r_j)}\ran=1$. On
the other hand, if particles $i$ and $j$ are not connected,
$|\r_i-\r_j|$  can assume any value and the above argument does not
apply. However, if particles $i$ and $j$ are not connected, we can
write
\begin{eqnarray}
\lan e^{-i\k\cdot(\r_i-\r_j)}\ran
=\frac{1}{N}\int\!d^3\r\,e^{-i\k\cdot\r}\,\rho h_{ij}(\r)
\label{meikr}
\end{eqnarray}
where $\rho= N/V$, $h_{ij}(\r) + 1 = g_{ij}(\r)$
and $(1/V)g_{ij}(\r)$ gives the probability 
density of finding the particle $i$ in $\r$, given the particle $j$ in the 
origin. 
We have used the fact that the Fourier transform of $1$ is zero
for any wave vector different from zero. The correlation function
$h_{ij}(\r)$ decays to zero at a finite distance,
so that (\ref{meikr}) is of order $O(1/N)$. As there are at most
$N^2$ disconnected pairs, the second term in Eq.\ (\ref{chi4sum}) is
of order $O(1/N)$, so that it can be neglected in the thermodynamic
limit. Finally, we have
$$
\lim_{k\to
0}\lim_{t\to\infty}\chi_4(k,t)=\frac{1}{N}\sum_{i,j=1}^N\,\left[
\gamma_{ij} \right]= \mbox{mean cluster size}.
$$
In the inset of Fig. \ref{fig5} the asymptotic value
$\chi_{as}(k_{min},\phi)=\lim_{t\to\infty}\chi_4(k_{min},t)$ is plotted
as a function of $(\phi_c-\phi)$
together with the mean cluster size.
We find that, as the percolation threshold is approached from below,
$\chi_{as}(k_{min},\phi)$ shows a power law behavior
with an exponent which, within the numerical accuracy, is in agreement
with the value of the exponent $\gamma$ of the mean cluster size.

\begin{figure}
\begin{center}
\includegraphics[width=8cm]{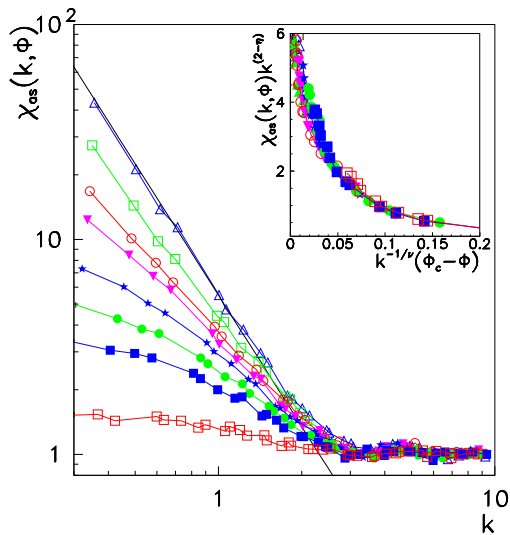}
\caption{{\bf Main frame}: $\chi_{as}(k,\phi_c)$ as a function of
the wave vector $k$ and, from bottom to top, $\phi= 0.02$, $0.05$,
$0.06$, $0.07$, $0.08$, $0.085$, $0.09$, $0.095$. At $\phi_c$ the
data are fitted with $\sim (k)^{\eta-2}$ (full line), with $2-\eta=
2.03\pm0.02$. {\bf Inset}: Scaling plot of $\chi_{as}(k,\phi)
k^{2-\eta}$ as a function of $k^{-1/\nu} (\phi_c -\phi)$.}
\label{fig7}
\end{center}
\vspace{-0.5cm}
\end{figure}

Using a standard scaling argument \cite{stauffer}, we can easily predict the
behavior of $\chi_{as}(k,\phi)$ for small $k$ close to $\phi_c$.
Since $k$ scales as the inverse of a length, $\chi_{as}(k,\phi) =
k^{\eta-2}f(k\xi)$, where $\xi \sim(\phi_c -\phi)^{-\nu}$,  $2-\eta=
\gamma /\nu$ and  $f(z)\sim z^{\gamma/\nu}$ for small values $z$ and
$f(z)\sim const$ for large values of $z$. Fig. \ref{fig7} shows that
the numerical data strongly support this scaling behavior.

In conclusion according to these results the percolation critical
exponents in the sol-gel transition could be measured via the
fluctuations of the self intermediate scattering functions by means
of light scattering. Moreover our results confirm that one key difference
between irreversible gelation due to chemical bonds and
supercooled liquids close to the glass transition is that in
irreversible gelation the heterogeneities have a static nature
(clusters). These clusters, on the other hand, affect the dynamics
and as a consequence the dynamic transition coincides with the
static transition, characterized by the divergence of a static
correlation length (linear size of the clusters). In this respect
the dynamical slowing down here is similar to the critical slowing
down found close to a second order critical point, with the
relaxation time diverging with a power law in the control
parameters. Interestingly the behavior of the dynamical susceptibility is very
similar to the one observed in spin glass models
\cite{lattice-gas}, where for long times it tends to the non linear static
susceptibility, which
diverges at the spin glass critical point. In fact, also in this
case the dynamic transition is connected to the divergence of a
static length.

What can we expect in colloidal gelation, where the bonds are
not quenched and have a finite lifetime? In principle, due to finite
bond lifetime, the clusters are not permanent anymore, consequently
we expect a structural arrest with a dynamical susceptibility of the
type found in glass forming liquid. This is in fact found in
experimental investigations of colloidal suspension
\cite{cipelletti} and in some molecular dynamic simulations
\cite{physicaa}. Interestingly enough, also in some spin glass
models, by introducing interactions with finite lifetime, there is
no divergence of the static susceptibility and one recovers the
behavior of the dynamic susceptibility which is typical  of
supercooled liquids \cite{lattice-gas}.

This work has been partially supported by the Marie Curie Fellowship
HPMF-CI2002-01945 and Reintegration Grant MERG-CT-2004-012867,
EU Network Number MRTN-CT-2003-504712, MIUR-PRIN 2004, MIUR-FIRB 2001,
CRdC-AMRA, INFM-PCI.

\end{document}